\documentclass[useAMS,usenatbib]{mn2e}
\usepackage{graphics}

\newcommand{\ex}{\mbox{$E(\mathrm{H}\beta - \mathrm{H}\alpha)$}}
\newcommand{\Teff}{\mbox{$T_{\mathrm {eff}}$}}
\newcommand{\Tbb}{\mbox{$T_{\mathrm {bb}}$}}
\newcommand{\Mz}{\mbox{$M_{\mathrm {ZAMS}}$}}
\begin{document}
   \title[The symmetric dust shell and the central star of
   NGC\,6537]
   {The symmetric dust shell and the central star of the 
   bipolar planetary nebula NGC\,6537
   \thanks{Based on observations with European Southern Observatory,
 Very Large Telescope with an instrument, ISAAC and NAOS-CONICA
(the proposal numbers: 65.D-0395A, 72.D-0766A).}
\thanks{ Based on observations made with the NASA/ESA Hubble Space
Telescope, obtained from the ESA/ECF Data Archive. HST is operated by
the Association of Universities for Research in Astronomy, Inc., under
NASA contract NASA 5-26555. These observations are associated with
program no. 8345.} 
          }

\author[M. Matsuura et al.]
{M.~Matsuura$^{1}$, 
A.A.~Zijlstra$^{1}$,
M.D.~Gray$^{1}$,
F.J.~Molster$^{2}$,
L.B.F.M.~Waters$^{3,4}$ \\
$^{1}$School of Physics and Astronomy, University of Manchester, Sackville Street, P.O. Box 88, Manchester M60 1QD, UK \\
$^{2}$ESTEC, European Space Agency, Keplerlaan 1, 2201 AZ Noordwijk, The
       Netherlands \\
$^{3}$Astronomical Institute `Anton Pannekoek', University of Amsterdam,
       Kruislaan 403, 1098 SJ, Amsterdam, 
       The Netherlands \\
$^{4}$Instituut voor Sterrenkunde, Katholieke Universiteit Leuven, 
       Celestijnenlaan 200B, 3001 Heverlee, Belgium \\
}
\date{Accepted. Received; in original form }
\pagerange{\pageref{firstpage}--\pageref{lastpage}} \pubyear{2005}

\maketitle
\label{firstpage}

\begin{abstract}
  We present high-resolution images of the strongly bipolar planetary
  nebula NGC\,6537, obtained with {\sc Hubble Space Telescope} and with the infrared
  adaptive optics system on the {\sc Very Large Telescope}.  The central star is
  detected for the first time. Using the multi-band photometry and
  constraints from the dynamical age of the nebula, we derive a
  temperature in the range 1.5--2.5$\times10^5\,$K, a luminosity $\sim
  10^3\,\rm L_\odot$, and a core mass $M_c \approx 0.7$--$0.9\,\rm M_\odot$.
  The progenitor mass is probably in the range $M_i = 3\hbox{--}7\,\rm
  M_\odot$.  The extinction map shows a largely symmetric, and compact dust 
  structure, which is most likely a shell, located at the neck of the bipolar flow, only
  2--4 arcsec from the star. The dust shell traces a short-lived phase
  of very high mass loss at the end of the AGB.  The dynamical age of
  the shell and bipolar lobes are very similar but the morphologies
  are very different. The data suggests that the mass loss during the
  ejection of the compact shell was largely spherically symmetric, and
  the pronounced bipolarity formed afterwards. The dynamical ages of
  the bipolar lobes and dust shell are similar, which is consistent
  with suggestions that bipolar structures form in a run-away event at
  the very last stages of the AGB mass loss.  The inner edge of the
  dust shell is ionized, and PAH emission is seen just outside the
  ionized gas.  We associate the PAH emission with the
  photo-dissociation region of the molecular shell.
\end{abstract}

%

\section{Introduction}

 At the end of the asymptotic giant branch (AGB) phase, stars
experience high mass loss rates
($10^{-7}$--$10^{-6}$\,$M_{\sun}$\,yr$^{-1}$), which may increase by a
further order of magnitude during helium shell flashes
\citep{Habing96}. The high mass loss terminates once the hydrogen
envelope of the star is almost fully removed.  During the subsequent
post-AGB phase, the temperature of the central star increases, and
eventually the surrounding ejected material is ionised, and detected
as a planetary nebula (PN).

During the so-called superwind, the objects suffer high
self-extinction from dust which forms in the ejecta; the absorbed
radiation is re-emitted at infrared wavelengths.  The IRAS photometric
observations show that the infrared colour temperature decreases from
1000\,K in AGB stars to a few hundred Kelvin in post-AGB stars and PNe
\citep{vanderVeen88, Zijlstra01}: as the star evolves, the
circumstellar shell expands and the temperature of the circumstellar
shell decreases.

Planetary nebulae commonly show non-spherically symmetric
morphologies, often elliptical or bipolar.  These may trace the
interaction of a fast wind ($> 1000\rm \,km\,s^{-1}$) from the central
star of the PN, with the slow wind (20--30\,km\,s$^{-1}$) dating from
the AGB \citep{Kwok82}.  The expansion velocity of the AGB shell is
20--30\,km\,s$^{-1}$ while the fast wind from the PN is more than
1000\,km\,s$^{-1}$. This interaction of the winds can enhance shapes
which are already present in the original AGB wind.  However, it is
still not clear how the original shape is created.  There is evidence
that bipolar PNe tend to have higher mass progenitors
\citep{Kastner96}, from helium and nitrogen enrichments characteristic
of type I PN \citep{Peimbert78}.  These have likely progenitor masses
of $> 2.5\,\rm M_{\sun}$.  Magnetic fields have been suggested, or a
binary companion may give rise to a circumbinary dusty disk
\citep{Balick02}.  A warped disk has been suggested to give rise to
multipolar PNe \citep{Icke03}. Some AGB stars show a cold detached
dust shell. Infrared images have resolves some of these shells, which
appear not to show deviations from symmetry \citep{Izumiura96,
Olofsson00}.


Here, we report high spatial resolution images of the PN NGC\,6537,
which shows one of the most pronounced bipolar outflows seen among
PNe. It is a type-I PN, which is over-abundant nitrogen and helium
\citep[c.f.][]{Pottasch00b}. The central star of NGC~6537 has not
previously been detected \citep{Pottasch00a}.  We use HST images to
create an extinction map: this traces the dust at far higher
resolution than can be achieved using far-infrared dust emission. We
find that the extinction inside this nebula is not uniform, but is
concentrated in a compact dust shell.  The central star is detected
for the first time, only 0.5 arcsec away from the brightest part of
the nebula.  
Using the photometry in the optical HST images and the
infrared adaptive optics image, the luminosity of the central star is
constrained.

\section{Observations}

We observed NGC\,6537 with the Adaptive Optics (AO) system, NAOS,
and the infrared camera CONICA, on the Very Large Telescope (VLT) on
the 6th of March 2004 (in UT).  The exposure time was 28\,minutes.  We
used the S27 camera with a pixel scale of 27 milliarcsec. The Ks-band
filter was used.  The dichroic for AO was VIS, i.e., the visual camera
was used.  The AO reference star was S\,300100217909 ($V$=14.8\,mag)
from the Guide Star Catalogue, which is approximately 22\,arcsec away
from NGC\,6537's central star.  The approximate spatial resolution of
the final image is 0.22\,arcsec (RA orientation) and 0.16\,arcsec (Dec
orientation), estimated from nearby stars; but the spatial resolution
varies slightly even within the image.  The weather was clear.  The
internal calibration lamp was used for flat fielding. The jittering
technique was used to minimize the effect of hot pixels.  The sky
background was estimated from a medium of jittered frames with
slightly different positions.

\begin{table}
\begin{caption}
{The log of HST observations. 
}\label{table-hst}
\end{caption}
\begin{tabular}{lllllll}\hline\hline 
Filter & Wave  & BW  & CS &$T_{\rm{ex}}$ & Lines\\ 
& [$\AA$] &[$\AA$]\\ \hline
F673N  &  6732  & 30  & yes & 400sec$\times$1,500sec$\times$1  & [S\small{II}]\\
F631N  &  6306  & 13  & yes & 400sec$\times$3                  & [O\small{I}]\\
F658N  &  6591  & 30  & no  & 400sec$\times$3                  & [N\small{II}]\\
F656N  &  6564  & 54  & no  & 400sec$\times$3,40sec$\times$1   & H$\alpha$\\
F547M  &  5483  & 205 & yes & 40sec                            & \\
F502N  &  5013  & 48  & no  & 300sec$\times$2, 400sec$\times$1 & [O\small{III}]\\
F487N  &  4865  & 32  & no  & 500sec & H$\beta$ \\
F469N  &  4695  & 17  & no  & 400sec & [He\small{II}]\\
\hline
\end{tabular}
Wave: wavelength \\
BW: band width (for photometric calibration; not equivalent with FWHM) \\
CS: detection of the central star\\
$T_{\rm{ex}}$: exposure time per frame (some bands have multiple
exposures)\\
Line: major lines inside the filter\\
\end{table}

NGC\,6537 was observed with the Hubble Space Telescope (HST) in the programme
6502 (P.I. B. Balick).  In total, HST images were obtained with 8 different
filters (Table\,\ref{table-hst}).  We retrieved the images from data archives
at the Space Telescope - European Coordinating Facility (ST-ECF).  The data
were taken on the 12th of September 1997 with the Wide Field Planetary Camera
2 (WFPC2), which has four 800$\times$800 pixel detectors. The main region of
interest for this paper, close to the centre, is covered with the Wide Field
Cameras which have a 0.1\,arcsec pixel scale.  The spatial resolution is
limited by the pixel scale, as the point-spread function is under-sampled, but
slightly better than that because multiple images were taken for a single band
(with the exception of F487N and F547M).  We use the pipeline-reduced data.
The PHOTFLAM parameter is used for the flux calibration.  Spike noise due to
cosmic rays was removed using the exposures with slightly different positions
if multiple images were obtained within the same filter, exceptions being
F487N and F547M which have a single exposure for each.  For F487N, we removed
the cosmic ray noise by taking the medium over nearby regions.

 Our PAH image was obtained with ISAAC on the VLT.  The data were
acquired on 13th of August, 2000 in service mode.  We used the {\it
eclipse} and IDL packages for data reduction.  Images were obtained
with two filters: NB$_-$3.28 (PAH band) and NB$_-$3.21 (continuum).
Total exposure time was 7\,min for each band.  Optical seeing was
0.56--0.83 and 0.47--0.66\, and the measured infrared seeing was
0.53--0.61 and 0.47--0.54\,arcsec (FWHM), respectively, which was
measured from the nearby stars.  After the flux calibration with
reference to a standard star HR~7446, the continuum was subtracted,
and a PAH image was created.

\section{ Results : the central region of NGC\,6537}


\begin{figure}
\centering
\resizebox{\hsize}{!}{\includegraphics*{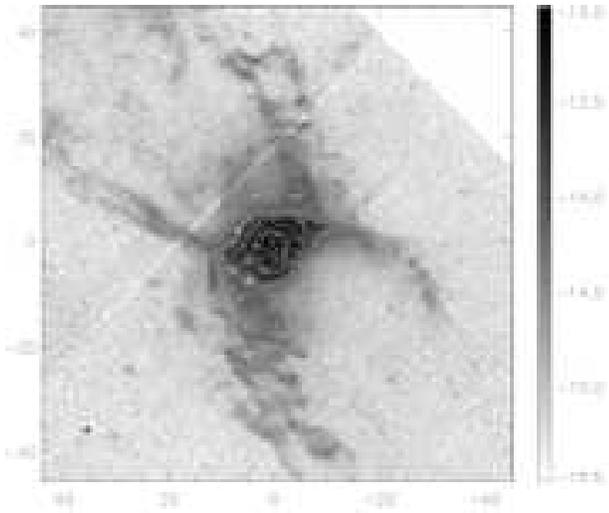}}
\caption{The entire field of the planetary nebula NGC\.6537 in H$\alpha$.
North is top and east is left. The x-axis is the RA offset in arcsec and 
y-axis is the Dec offset in arcsec, for all of the images in this paper.
The unit is erg\,cm$^{-2}$s$^{-1}$arcsec$^{-2}$, 
contours show log ($I_{\nu}$)=$-14, -13.6, -13.3, -13, -12.6, -12.3, -12$.
}
\label{Fig-Ha-large}
\end{figure}

\begin{figure}
\centering
\resizebox{\hsize}{!}{\includegraphics*[77,374][474,730]{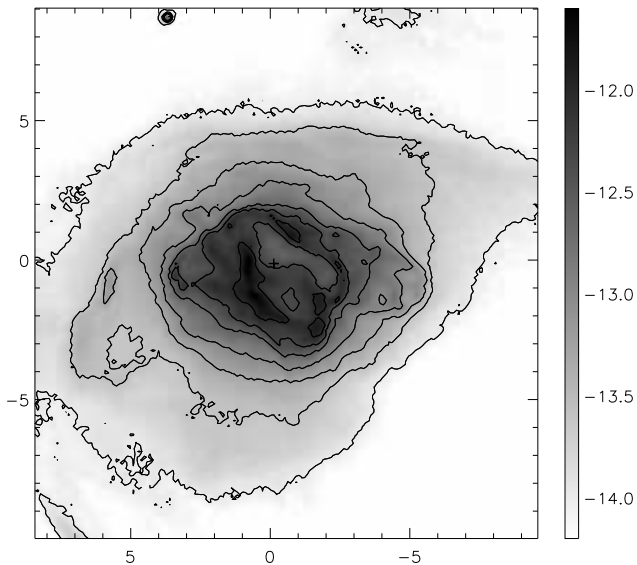}}
\caption{
The central region of H$\alpha$ ($I_{\nu}$) image
in the unit of erg\,cm$^{-2}$s$^{-1}$arcsec$^{-2}$, 
contours show log ($I_{\nu}$)=$-14, -13.6, -13.3, -13, -12.6, -12.3, -12$.
(same as Fig.~\ref{Fig-Ha-large})}
\label{Fig-Ha}
\end{figure}

The entire field at H$\alpha$ is shown in Fig.\,\ref{Fig-Ha-large}.  It shows
an extended bipolar outflow, and a dense core with a size of
$\sim$5$\times$5\,arcsec$^2$ at the centre.  The bipolar lobes have the
appearance of an hourglass, seen close to edge-on; the north-western lobe is
slightly fainter and may be pointing away from us. At the faintest levels, the
nebula extends beyond the WFPC2 field of view. The lobes extend over 1
arcminute, and are oriented roughly NE-SW. The core consists of
bright arcs tracing a shell surrounding an irregular cavity
(Fig.~\ref{Fig-Ha}).  The cavity is between 2.5 and 4 arcsec in diameter, with
the long axis roughly (but not perfectly) aligned with the bipolar flow.  The
arcs have thickness of approximately 1 arcsec.  There is a much fainter
extended shell with a diameter of $6 \times 2.5$ arcsec, with the major axis
oriented EW. Apart from the cavity elongation, none of the structures shows an
obvious alignment with the bipolar axis.

\begin{figure}
\centering
\resizebox{\hsize}{!}{\includegraphics*[17,11][360,354]{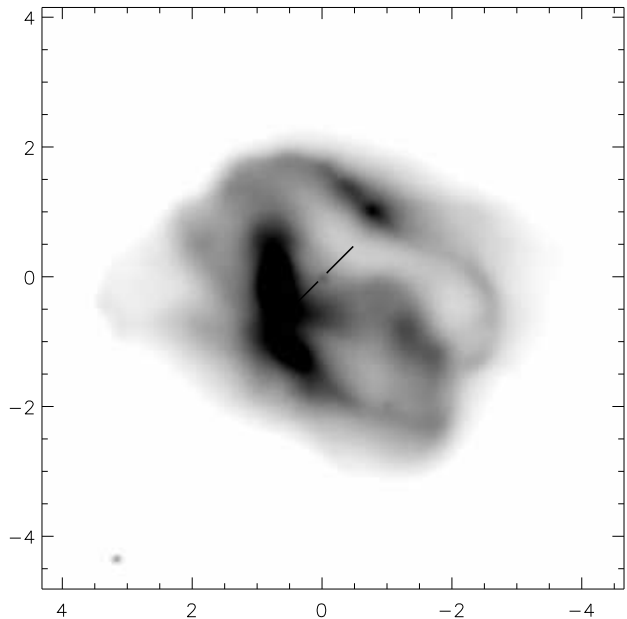}}
\caption{Ks-band image (in linear scale). The lines show the position of the central star.}
\label{Fig-K}
\end{figure}

The Ks-band image is shown in Fig.\,\ref{Fig-K}.  Several arcs are
detected, which are the same as those found in the central region of
the H$\alpha$ image, as discussed later.  The dominant source of
emission in the Ks-band is likely to be free-free and bound-free
continuum emission from the ionized gas. Only regions of higher
emission measure in the ionised region are detected at the
Ks-band. The much fainter bipolar wings are not detected.

\subsection{ Detection of the central star }

\begin{table}
\begin{caption}
{Measured fluxes of the central star. The error includes
only the measurement error of aperture photometry.
Systematic errors, such as contribution of nebula into the aperture
or the error of calibration star (for Ks band) are not included.
}\label{table-mag}
\end{caption}
\begin{tabular}{llllll}\hline\hline 
Filter & Wavelength [$\AA$] & Magnitude & Flux [Jy]  \\ \hline
F673N  &  6732   & 21.74$\pm$0.54 & $[11.0\pm7.2]\times10^{-6}$ \\
F631N  &  6306   & 21.34$\pm$0.33 & $[14.2\pm5.1]\times10^{-6}$ \\
F547M  &  5483   & 21.56$\pm$0.22 & $[ 8.6\pm2.0]\times10^{-6}$ \\
Ks     & 2.18$\times 10^4$   & 18.63$\pm$0.61 & $26\times10^{-6}$ \\
\hline
\end{tabular}
\end{table}


\begin{figure}
\centering
\resizebox{\hsize}{!}{\includegraphics*[71,375][417,774]{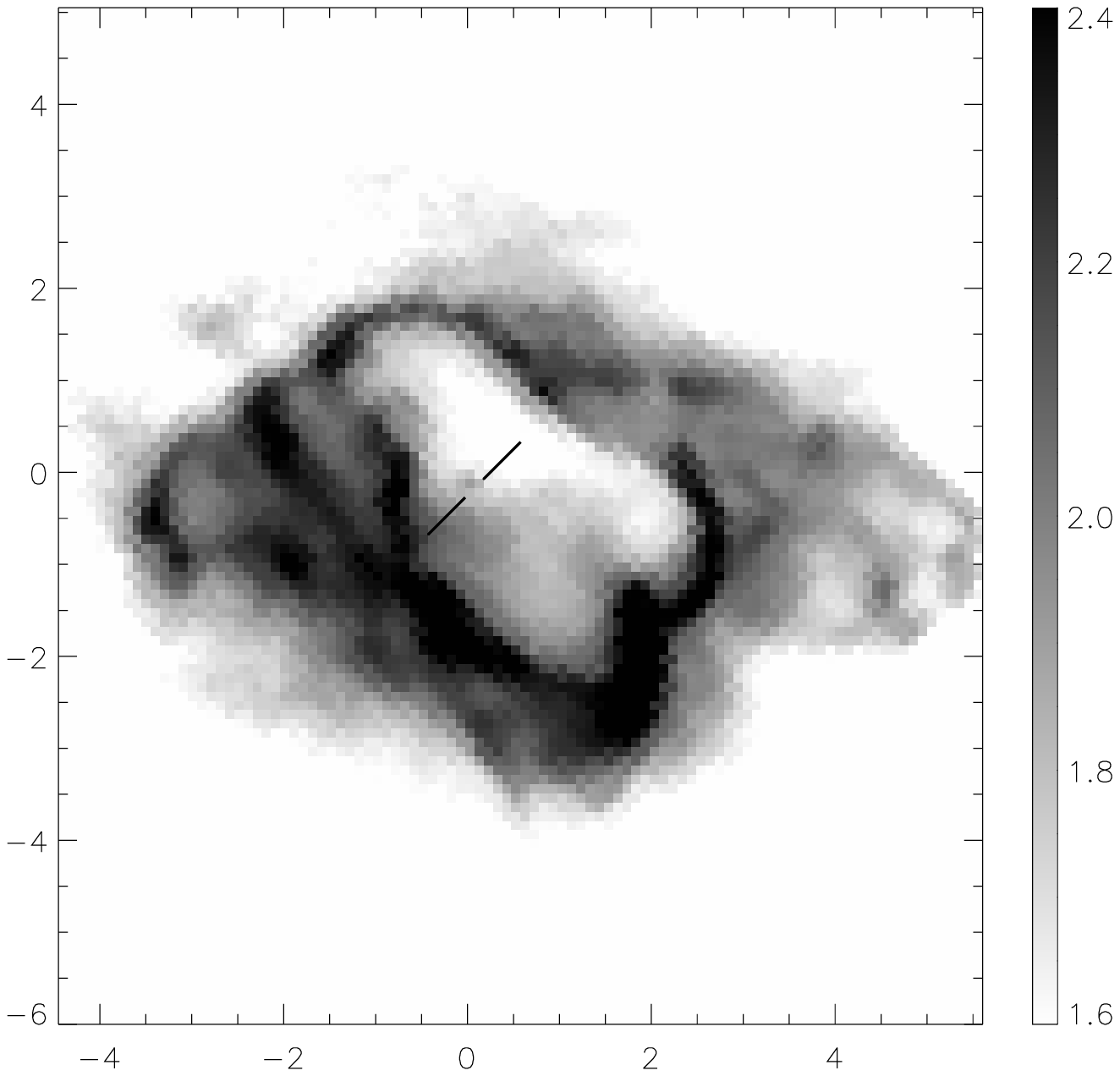}}
\caption{HST F673 band image (in logarithmic scale). The lines show the
position of the central star.
}
\label{Fig-F673}
\end{figure}

\begin{figure}
\centering
\resizebox{\hsize}{!}{\includegraphics*[71,375][417,774]{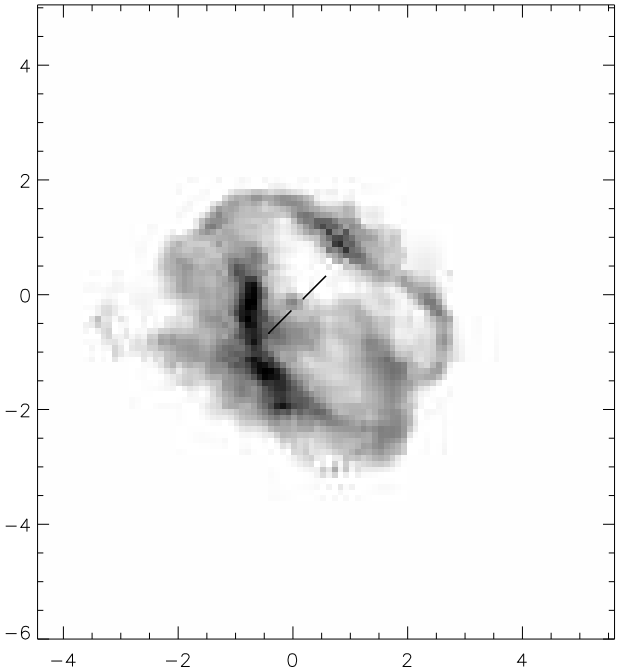}}
\caption{HST F547M band image (in logarithmic scale). The lines show the
position of the central star.
}
\label{Fig-F547}
\end{figure}

The star is clearly detected, separated from the extended nebula in the high
spatial resolution images with HST and NACO.  The coordinates of the newly
discovered central star are RA=18\fh05\fm13\fs03,
Dec=$-$19\fdg50\farcm34\farcs4 (J2000).

The Ks magnitude of the central star is $18.6\pm0.6$~mag, which is derived
from aperture photometry with a radius of 5 pixels.  The magnitude is
referenced to nearby stars in the same frame: 2MASS~18051346$-$1950277
(Ks=13.382$\pm$0.166~mag) and 2MASS~18051378$-$1950407
(Ks=12.674$\pm$0.040~mag).  The uncertainty of 0.61\,mag is the measurement
error only: the uncertainty from the magnitude calibration is not included.
There might be some flux contribution from the nebula within the aperture, which may
not have been removed completely. The background level (including the
contribution of nebula) is estimated from circular with inner radius
of 150 \% and outer radius of 200 \% of the apeature size.

The central star is also detected in three HST filters, two of which are shown
in Figs.\,\ref{Fig-F673} and \ref{Fig-F547}.  The results of aperture photometry with a radius of
2 pixel are listed in Table\,\ref{table-mag}.  The largest uncertainty of the
magnitudes is due to the contribution of nebula and the flux missed outside of
the radius. These systematic errors are not counted in these errors.  
F547M is
the only medium-band filter  and this filter is
sensitive to continuum.  In addition, the central star is detected in
two narrow band filters.
The remaining five filters which were used for the
observations did not give detections. The star is too faint to be detected in
the narrow blue filters (F487N, F469N and F502N; Fig.~\ref{Fig-Hb}), due to extinction, while the
F656N and F658N filters contain strong ionised lines within the filter
transmission (H$\alpha$ and [N{\small{II}}]), and the contamination of these
ionized lines from the nebula prevents detection of the star
(Fig.~\ref{Fig-Ha}).  All three HST filters where the central star was
detected have only weak atomic lines within their wavelength coverage.

\citet{Pottasch00a} reported a non-detection of the central star, using the
same HST data.  The under-sampling of the Wide Field Camera can make it
difficult to distinguish a faint star from spike noises due to the cosmic ray
hits, and this may be the reason for their non-detection.  However, comparison
with the well-sampled Ks-band images confirms the central star detection in
three images.

\citet{Kaler89} predicted that the central star of NGC\,6537 should be about
$V=19.67$\,mag, and \citet{Pottasch00a} estimate $V=22.4$\,mag.  We obtained
magnitudes in a `continuum band' at 5483\,\AA, which is very close to the
V-band. Assuming that the visual magnitude is the same as that in the F547M
band, 21.6$\pm$0.22\,mag and ignoring the systematic errors, our measured
magnitude is just between the predicted values of \citet{Pottasch00a} and
\citet{Kaler89}.

A detection of the central star has been suggested by \citet{Reay84}, as a
`stellar-like condensation to the east of centre.'  However, \citet{Gathier88}
and \citet{Pottasch00a} show that this detection is not a star but a
condensation in the nebula.  We agree with the latter view, since the nearby
condensed part of nebula is only $\sim0.7$ arcsec away from the star, and
could not have been resolved from the star at 2\,arcsec seeing \citep{Reay84}.

\subsection{The extinction map }

\begin{figure}
\centering
\resizebox{\hsize}{!}{\includegraphics*[77,374][474,750]{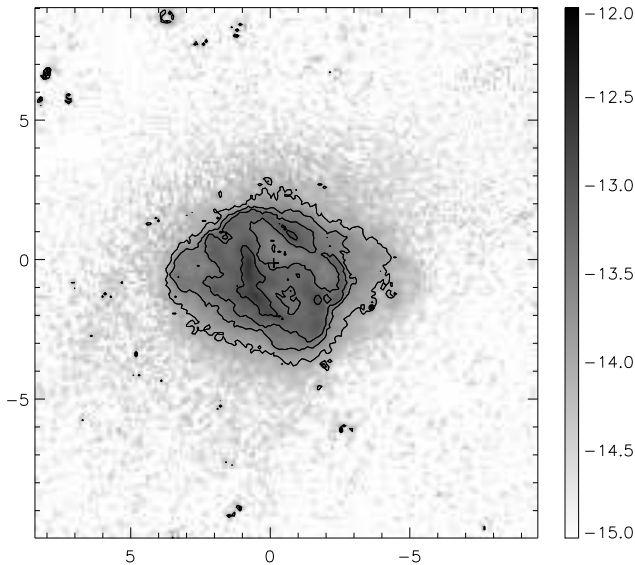}}
\caption{
The central region of H$\beta$ ($I_{\nu}$) image
Contour show log ($I_{\nu}$)=$-14, -13.6, -13.3, -13$.}
\label{Fig-Hb}
\end{figure}
\begin{figure}
\centering
\resizebox{\hsize}{!}{\includegraphics*[77,374][474,750]{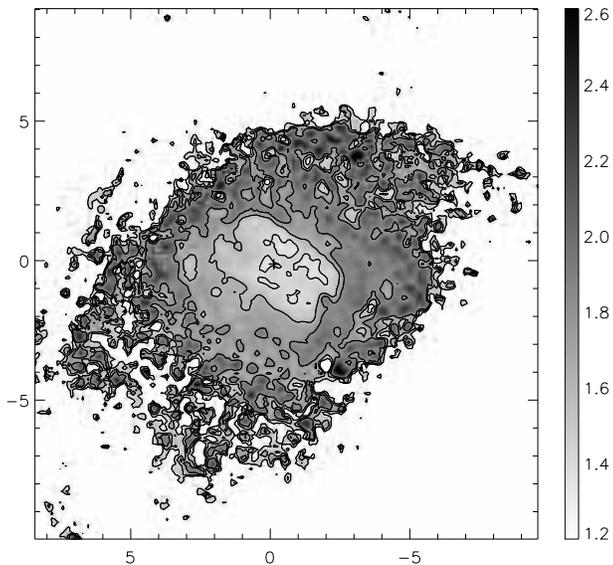}}
\caption{
The extinction map.
The contour lines show 1.4, 1.6, and 1.8~mag in \ex, and smoothed to
0.14~\,arcsec\, pixel$^{-1}$.
The central star is marked as a cross.
At the centre, the extinction is 1.4--1.5~mag, while the extinction is 
higher than 2~mag at $\sim$4~\,arcsec\,  from the central star.
White points inside the extinction map (e.g. RA offset $\sim-2$ and
Dec offset of $\sim-4$) are caused by cosmic ray hits
on the detector, in the H$\beta$ image.
}
\label{Fig-ex}
\end{figure}

The extinction map is estimated from HST H$\alpha$ (F656N) and H$\beta$
(F487N) images.  We assume an electron temperature of 15\,000\,K, and an
electron density of $1\times10^4$ \,cm$^{-3}$ \citep{Pottasch00b}.  The
H$\alpha$ and H$\beta$ intensity ratio of 2.79 (Case B) is used
\citep{Storey95}. The sensitivity is limited by H$\beta$ data
(Fig.~\ref{Fig-Hb}). The image is smoothed to 0.2 arcsec pixel scale.
This also removes any difference in resolution between H$\alpha$ and
H$\beta$.

Fig.~\ref{Fig-ex} shows the extinction map of the core region of NGC~6537.
The extinction within the core is non-uniform, showing the presence of
internal extinction within the PN.  At the centre, the extinction \ex\, is
1.4--1.5~mag, while it increases up to 2.2~mag at 4~\,arcsec\, away.  The
thickness of the high extinction region is about 2\,arcsec.  This high
extinction shell shows the presence of a detached dust shell, located far
inside the extended planetary nebula. The inner cavity of the dust shell shows
an elongation along the bipolar outflow which is towards the
north-east(=left-top in the figure), and the south-west(=right-bottom).  The
central star is located at the centre of the elongated dust shell.  There are
small `patch-like' scale structures inner region the shell within 2--3 arcsec
from the central star.

\citet{Cuesta95} suggested a reduction in extinction at the centre,
using their optical spectroscopy across the centre region.
We confirm this, and show the extinction peaks in a shell structure.

The average extinction towards NGC~6537 is reported as $E(B-V)$=1.23
\citep{Pottasch00a}, $A_V$=3.4 \citep{Cuesta95}, or $C$=1.79--2.16
\citep{Ashley88}.  These correspond to \ex=1.50, 1.33, 1.48--1.79, if the
interstellar extinction law \citep*{Cardelli89} and $A(V)/E(B-V)$=3.1 are
assumed.  These values of extinction were derived from hydrogen recombination
lines, presumably weighted towards the lines originating from high H$\alpha$
intensity region.  In our map, that region has $\ex = 1.5$~mag, which is
within previous measurement uncertainties.  The average \ex\, of the whole
region (defined as \ex\, higher than 0.2 and smaller than 3.0~mag) is 1.25,
and the median is 1.20~mag, which is below \ex\, at the high H$\alpha$ region.

\subsection{Hierarchy of H$\alpha$, PAHs and extinction}

\begin{figure}
\centering
\resizebox{\hsize}{!}{\includegraphics*[87,374][484,710]{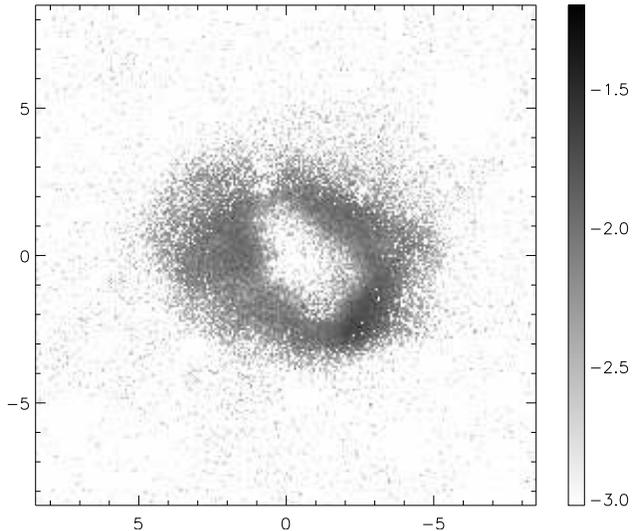}}
\caption{
PAH image in Jy~arcsec$^{-1}$.
}
\label{Fig-PAH}
\end{figure}

\begin{figure}
\centering
\resizebox{\hsize}{!}{\includegraphics*{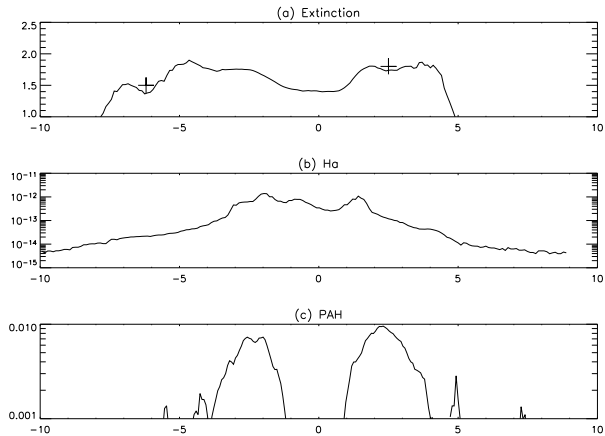}}
\caption{
A slice through the extinction map, H$\alpha$ map, and PAH image 
at constant RA  (minus is the south) across the central star.
Extinction is smoothed to 0.8\,arcsec pixel$^{-1}$ on image. Bumps at Dec
offset =$-6.2$ and 2.5 in the extinction map marked with crosses
are artificial (one large spike noise spread to
other pixels because of the smoothing).
}
\label{Fig-slice}
\end{figure}

Fig.\,\ref{Fig-PAH} show the PAH image of the central region. The PAH emission
is concentrated in an elongated shell. Comparing with the H$\alpha$
distribution (Fig.\,\ref{Fig-Ha}) shows that the arcs seen in the H$\alpha$
image are also emitting PAHs, but the shapes are not fully coincident: the PAH
distribution traces the outer edge of the optical arcs.  Slices through the
H$\alpha$, PAH and extinction images are plotted in Fig.\,\ref{Fig-slice}.
This shows a radial hierarchy in the shells and rings: the H$\alpha$ rings are
found in the innermost region, and PAHs are found at and outside of the
H$\alpha$ region, where the extinction is $E(B-V)$$\sim$1.8~mag, slightly
higher than the extinction towards the central star.  The extinction increases
further outside of the PAH emitting region. 

To first approximation, the dense H$\alpha$ emission traces the inner
edge of the dust shell, while the PAH emission could be located in a
photo-dissociation region. However, confirmation of such a model would
require detection and mapping of molecular emission.


\section{Discussion}

\subsection{The central star}\label{sec-central-star}

The magnitudes of the central star allow us to estimate the luminosity of the
central star, and to discuss its evolutionary stage.  To derive the
luminosity, the effective temperature and extinction towards the centre are
needed.

For the extinction towards the central star we take the central extinction
measured from H$\alpha$/H$\beta$. This assumes that the ionized region is
located within the dust shell. The hierarchy discussed above supports this
distribution. We ignore back-scattering which may make the observed extinction
slightly higher. If this were important, the dense H$\alpha$ filament located
very close to the position of the star could be expected to show lower
extinction. However, the filament shows identical extinction to the general
inner region.

We estimate the Zanstra temperature using our visual magnitude of the
central star and
mean extinction, together with the total flux of H$\beta$ and He 4861 $\AA$\,
\citep{Reay84}.  The Zanstra temperature estimated from helium is 340,000\,K.
The estimate of \citet{Pottasch00a} shows a higher Zanstra temperature of
500\,000\,K.  Since we can not estimate the systematic error
in our measurements of visual magnitude, 
this higher temperature could not be ruled out totally at this point.

Given the very high temperature, the various uncertainties on the Zanstra
temperatures need to be considered.  Some cause the He\small{II} Zanstra
temperatures to be too low: high optical depth, internal dust extinction,
helium overabundance \citep{Gruenwald00}. A binary companion could make the
star appear brighter which would also lead to a lower Zanstra temperature. An
excess of stellar photons in the extreme ultraviolet, shortward of the He
ionization edge, can cause the Zanstra temperature to be overestimated.
\citet{Henry86} discuss the importance of composition: He-poor atmosphere
cause the HeII Zanstra temperature to be overestimated, however this effect
diminishes at the highest temperatures and is perhaps less likely for a type-I
(He-rich) PN. \citet{Gabler91} mention that shocks within the fast wind can
increase the EUV emission and lead to overly high Zanstra temperatures.

\begin{figure}
\centering
\resizebox{\hsize}{!}{\includegraphics*{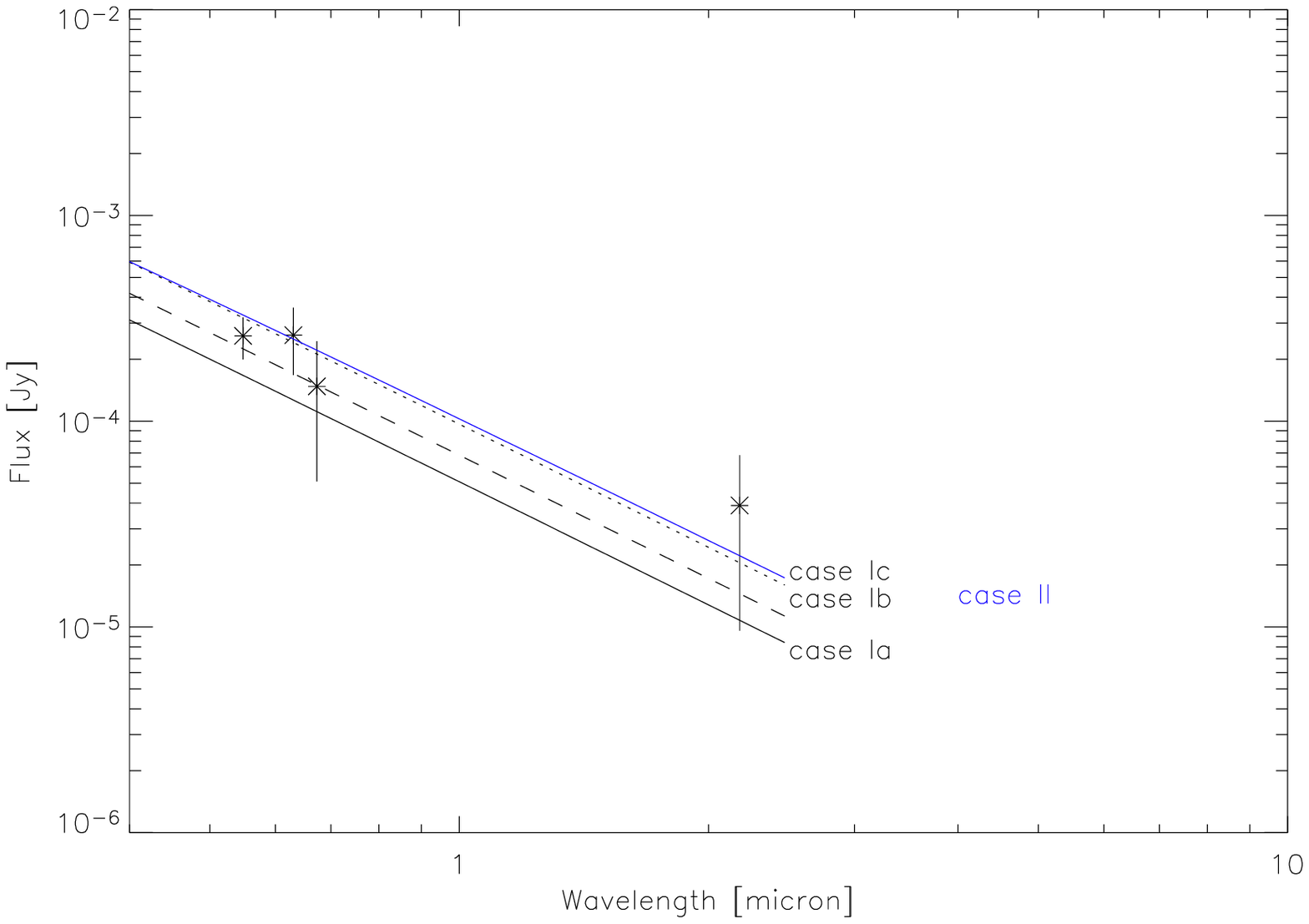}}
\caption{
The extinction-corrected flux of the central star, and the black body fits.
The parameters are listed in Table\,\ref{table-teff}.
Case Ic and case II have almost the same line, the dashed line 
is for case Ic, and the blue line is for case II.
}
\label{Fig-flux}
\end{figure}

{}From the previous section, the extinction towards the central star is
\ex=1.40, which is equivalent with $A(V)=3.71$.  The magnitudes for the four
bands where the star is detected, are corrected for this extinction. We fit the
resulting fluxes with a black body (Fig.\,\ref{Fig-flux}) to derive the
luminosity of the central star ($L_{\star}$) (Table~\ref{table-teff}).
Initially, parameters in \citet{Pottasch00a} are used; the black body
temperature is \Tbb=500\,000~K, the distance is $D_0$=2.4~kpc, and the stellar
radius is $r_0=1.16\times 10^{-2} R_{\sun}$.  With these parameters (case Ia),
the fit is within 1\,$\sigma$ in the optical but outside of this range at
F547M and F631N.

For the next step, we scale the blackbody by a factor $a^2$, i.e. the solid
angle is $\Omega \approx \pi (r_0/D_0 \times a)^2$.  The lowest flux within
the uncertainty (1 $\sigma$) is found at $a$=1.16 (case Ib) and highest flux
is found at $a$=1.38 (case Ic).  The luminosity is calculated with two
possibilities for each case, adopting the scaling factor for distance and for
radius, respectively.  We assume \Tbb\, is equivalent with effective
temperature \Teff.  Consequently, the luminosity range is
[7.5--10.1]$\times10^3$~$L_{\sun}$ (Table~\ref{table-teff}).

Finally, \Tbb\, is varied.  A lower estimate of the temperature of the central
star has been reported, and \citet{Casassus00} estimated 150\,000~K from the
ionization levels of the nebula.  The fitted result corresponds to case (II)
in Table~\ref{table-teff}.  The luminosity is an order of magnitude lower in
this case.  On the other hand, with our estimated Zanstra temperature of
340\,000~K, the luminosity range is [1.6--4.6]$\times10^3$~$L_{\sun}$
(Table~\ref{table-teff}).  Case IIIa is for the lower limits of the magnitudes
errors, and case IIIb is for the upper limits.  The black body lines of case
III are not plotted in Fig.\,\ref{Fig-flux}, because the lines overlay with
those of case Ib and Ic on this plotting scale.

\begin{figure*}
\centering
\resizebox{\hsize}{!}{\includegraphics*{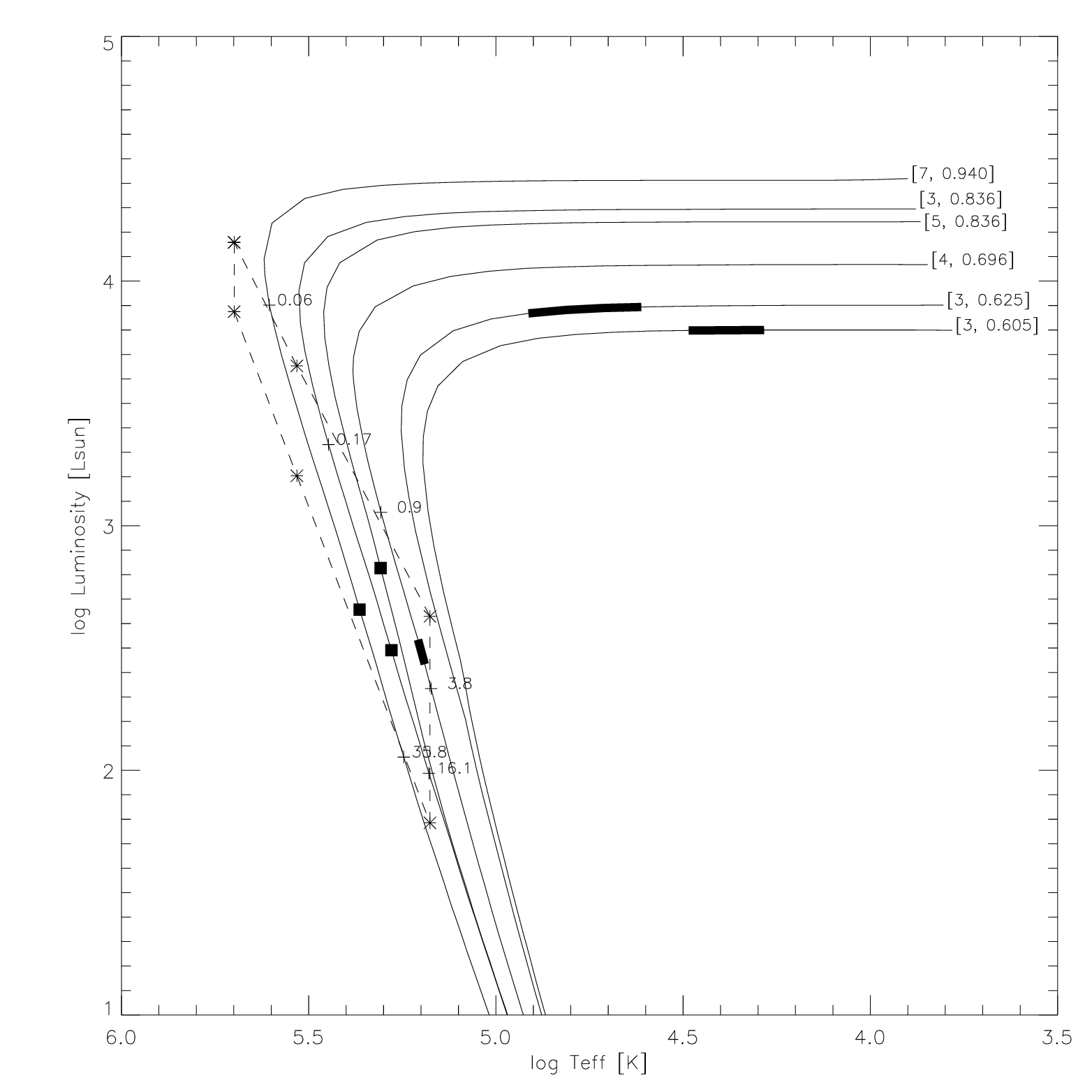}}
\caption{The estimated luminosity and \Teff\, of the central star of
NGC\,6537, plotted on the evolutionary tracks of post-AGB stars
\citet{Bloecker95}.  The dashed line encloses the estimated temperature and
luminosity range (the asterisks show the fit calculations).  The numbers in
the parenthesis are \Mz\, and the core mass at the tip of the AGB, in solar
units, for \citet{Bloecker95}'s models.  The numbers on the plus signs
show the age in $10^3$ years after the AGB phase.  The bold lines
on the evolutionary tracks indicate the age of 1000--2000 years after
the AGB phase, suggested by the dynamical age. This age range is estimated from the size of the dust shell.
$M_c \approx 0.7-0.9\,\rm M_\odot$ and \Mz\, of 3--7 $M_{\sun}$ is likely for NGC~6537.  }
\label{Fig-hr}
\end{figure*}

We plot these estimated $L_{\star}$ and \Teff\, (assuming \Tbb=\Teff) on the
HR diagram with a comparison with the theoretical evolutionary tracks for
solar abundances \citep{Bloecker95}.  Fig.~\ref{Fig-hr} shows that the central
star is within the highest luminosity as a PN and temperature range if \Teff\,
is 500\,000~K, whilst if \Teff\, is lower than that, the central star is
already in the white dwarf (WD) phase.  Different estimates for \Teff\ give
totally different luminosities.  Nevertheless, the models tend to coincide
with the higher mass tracks.

A further constraint on the stellar mass derives from the time scales for the
post-AGB evolution \citep{Gesicki00}. Higher-mass stars evolve much more
rapidly to higher temperature. All models accelerate when evolving through the
knee of the HR diagram, and evolution slows dramatically around
$10^3\,L_{\sun}$ for the highest mass tracks, and around $300\,L_{\sun}$ for
the 3\,M$_\odot$ track.  Indicative post-AGB ages for different tracks are
shown in Fig. \ref{Fig-hr}.

\citet{Cuesta95} reported that the central region of NGC\,6537 has an
expansion velocity of 18\,km\,s$^{-1}$. \citet{Corradi93} also found a similar
velocity. In previous discussion, we obtained distance of 0.9-2.4\,kpc. If we assume the distance of 1.5\,kpc, the dynamical age of the dust shell
should be about 1600\,years, assuming that the shell has been constantly
expanding with this speed and that the current radius is 4\,arcsec.
Models with core masses in the range 0.696 -- 0.943 \,M$_\odot$ gives a
possible fit for a luminosity in the range $2.5<\log L <3$. Lower-mass tracks
evolve too slowly before reaching the dashed region. The best fit track has a
core mass of $\approx 0.7--0.9$\,M$_\odot$ and progenitor mass of 3--7 \,M$_\odot$.

The comparison of ages and stellar parameters favours a lower
temperature for the central star of NGC\,6537, in the region
1.5--2.5$\times 10^5$\,K.  This is below the HeII Zanstra temperature,
suggesting the latter may have been overestimated.

The progenitor mass heavily depends on the mass lost during the AGB phase, and
is less well constrained than the core mass.  The relation between final and
initial mass may differ from the ones of the Bl\"ocker tracks. However, the
derived core mass is high for PNe \citep{Gesicki00} and is consistent with a
higher-mass progenitor. The nitrogen and helium enrichment of this type-I PN
\citep{Peimbert78} also suggests a high-mass progenitor
\citep[c.f.][]{Pottasch00b}.

\begin{table}
\begin{caption}
{
Fitting results to the fluxes.
}\label{table-teff}
\end{caption}
\begin{tabular}{llllll}\hline\hline 
       & \Tbb & $a$ & $L_{\star}$ \\
       & [K] & & [$L_{\sun}$] \\ \hline
case Ia   & 500\,000 & 1.00   & 7.5$\times10^3$ \\
case Ib   & 500\,000 & 1.16   & [7.5--10.1]$\times 10^3$ \\
case Ic   & 500\,000 & 1.38   & [7.5--14.4]$\times 10^3$ \\
case II   & 150\,000 & 2.64   & 61--426 \\
case IIIa & 340\,000 & 1.41   & [1.6--3.2]$\times 10^3$\\
case IIIa & 340\,000 & 1.68   & [1.6--4.6]$\times 10^3$\\
\hline
\end{tabular}
\end{table}

The bipolar lobes have velocities of $\sim 350\,\rm km\,s^{-1}$.  This gives
a similar expansion age to that of the core.  This suggest the
bipolar structure and the dust shell formed near-simultaneously,
and were present very soon after the onset of the post-AGB
evolution.

\subsection{Dust within NGC\,6537}
\label{dust}
\begin{figure}
\centering
\resizebox{\hsize}{!}{\includegraphics*{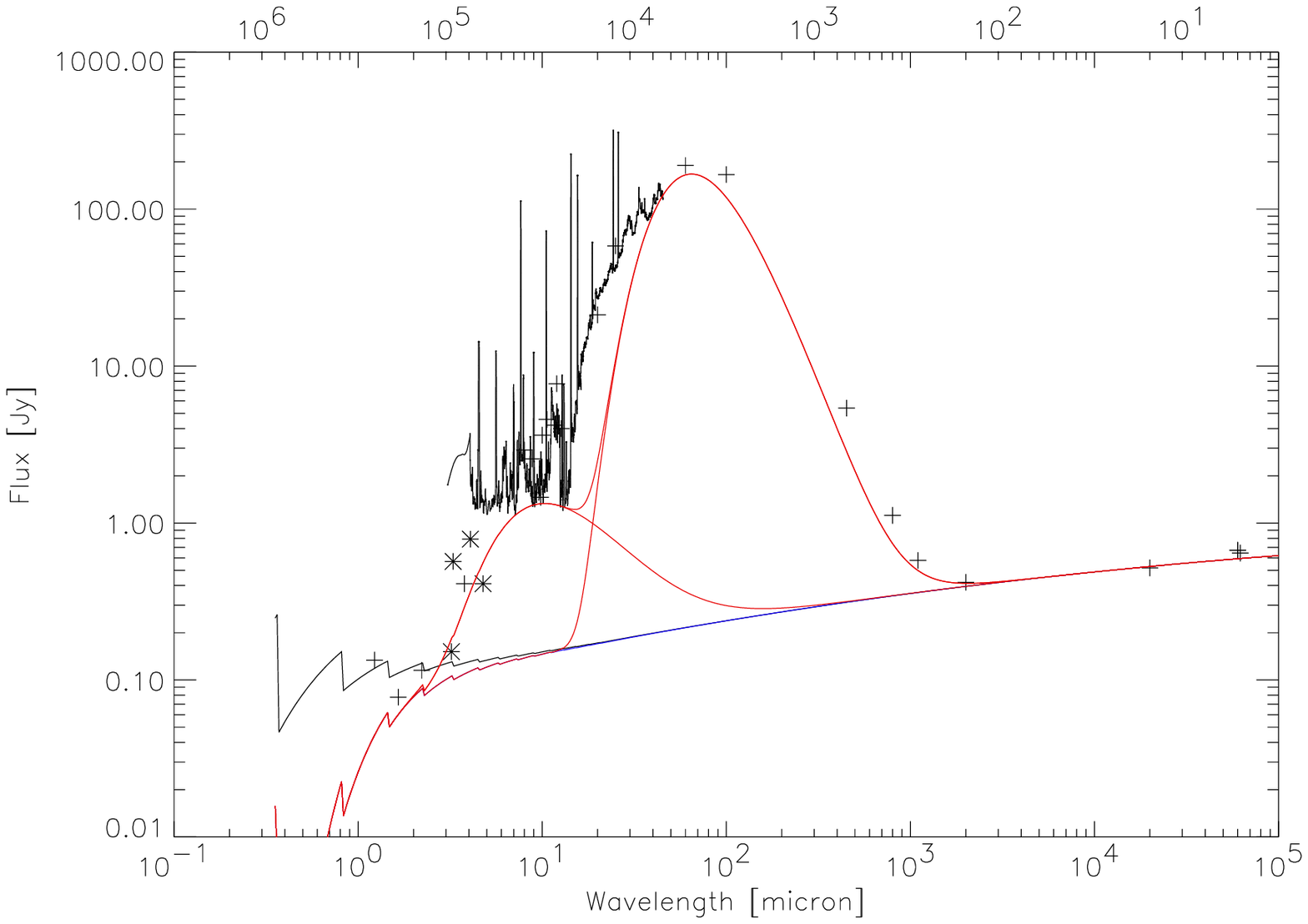}}
\caption{ The measured infrared to radio flux of the core of
NGC\,6537.  The data are from our ISAAC observations at L and M-bands,
ISO/SWS, IRAS, \citet{Whitelock85}, \citet{Phillips94},
\citet{Cohen80}, and \citet{Hoare92}.  The solid line is a total
of calculated
free-free and bound-free emission, 
scaled to the radio data.  
Two black body curves are shown (dashed lines) to emphasize the dust
excess on top of the extinction corrected free-free and bound-free
continuum.  }
\label{Fig-sed}
\end{figure}

The temperature of the circumstellar dust shell decreases as the circumstellar
shell expands and the central star evolves from AGB star to post-AGB star and
PN \citep{vanderVeen88}.  The cool far-infrared excess found in PNe is the
remnant of the large mass loss phase at the end of AGB phase.

The temperature of the dust shell is estimated from the central star
parameters and the distance, discussed in the previous section.  The high
extinction is mainly found within 2--4 \,arcsec from the star.  Ignoring the
projection effects, an inner radius of 2\,arcsec\, and outer radius of
4\,arcsec are assumed.  We can estimate the dust temperature $T_{\mathrm d}$
from $d_{\mathrm d} / R_{\star} = (\Teff/T_{\mathrm d})^2$, if the dust grains
are directly heated by the central star, where $d_{\mathrm d}$ is the distance
of a dust particle from the central star and $R_\star$ is the stellar radius.
For \Teff=3.4$\times 10^5$~K, scale $a=2.6 $ (Case IIIa in
Table\,\ref{table-teff}), $T_{\mathrm d}$ is 45\,K and 32\,K at 2\,arcsec, and
4\,arcsec\, in radius, respectively.  This can be compared to the dust 
condensation (and formation) temperature,
about 1\,000\,K for silicate. The dust grains formed much closer to the star,
during the AGB, and the grains have moved outwards.

 Infrared spectra (Fig.\ref{Fig-sed}) show two peaks in the
dust excess.  The colder shell has a peak at about 50--60\,$\mu$m,
which corresponds to temperature of about 50\,K. This is consistent
with the above estimated temperature from the dust radius.  The hotter
dust peaks around 10$\mu$m, with a temperature around 300\,K. This
hotter dust cannot be located much closer to the star, as the optical
and infrared images indicate a central cavity. Instead this emission must
arise for a component which reaches higher temperatures, either
because it consists of smaller grains, or receiving additional
heating. This component has a much smaller mass than the cold dust.


The extinction allows us to map the cold dust at much higher resolution than
would otherwise be possible. Imaging thermal emission of cold dust requires
far-infrared space telescopes such as ISO and Spitzer; their small apertures
are ``resolution-challenged'' \citep{Rieke04}; e.g. the 6-arcsec resolution at
24$\mu$m for Spitzer would leave the core of NGC\,6537 unresolved. In
contrast, our extinction map has a resolution of 0.14 arcsec per pixel. As the
dust shell closest to the core was ejected last, this gives a unique handle on
the morphology of the final AGB wind, and helps constrain the origin of the
bipolar morphology.

The extinction map is far more symmetric than either the HST or the infrared
images. The structure delineated by the inner cavity shows an elongation
approximately (but not perfectly) along the bipolar axis, with an axial ratio
of 1.5:1.  For comparison, the optical image shows an elongation of the full
structure (at the faintest emission levels) of approximately 4:1. A large
amplification has clearly taken place. The  structure shows that the mass 
loss on the AGB was much more spherically symmetric than the current PN would
suggest.

The shell is relatively thick, with an outer radius at least twice the
inner radius. The optical emission appears embedded in the dust shell,
and so the extinction map measures the extinction from the front shell
only (this is the reason we can apply the extinction map for the
central star). The observed ratio of peak extinction over central
extinction (in mag) of $1.4$, would require an outer radius of the
dust shell of 6 arcsec, for an observed inner radius of 3 arcsec and
uniform density.  However, the extinction includes an ISM foreground
component,  and the observed extinction ratio is therefore a lower
limit. A higher ratio corresponds to a smaller outer radius.

The dust shell may well extend further out than derived from the extinction
map: the optical H$\beta$ flux becomes too faint to accurately determine
extinction values. But the estimate above makes is likely that the dense part
of the shell does not extend much beyond the limits of the map.  The AGB wind
will extend much further as a $1/r^2$ envelope, as suggested by the large
extend of the optical emission. But an extended $1/r^2$ envelope would not
give rise to a large difference in extinction over the (small) central hole.
We conclude that inner dust shell is distinct from the outer AGB wind, either
tracing a short lived ($<10^4$\,yr) event of very high mass loss, and/or
swept-up gas.

\begin{figure}
\centering
\resizebox{\hsize}{!}{\includegraphics*{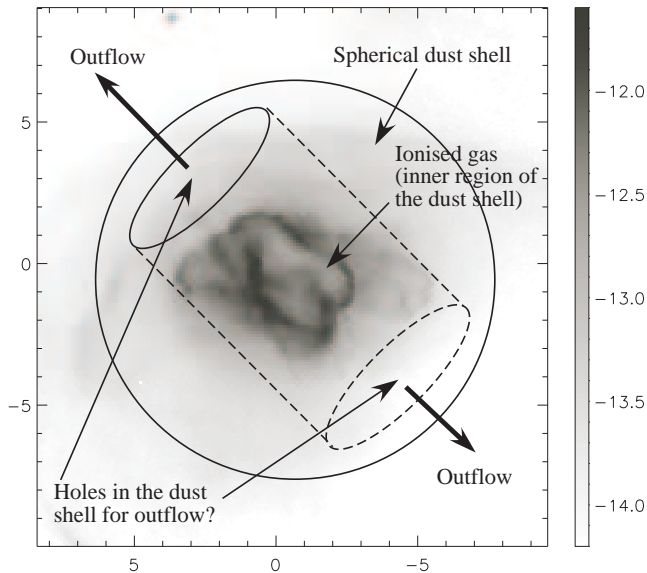}}
\caption{Schematic view of shell. Circumstellar shell may be elongated
along the holes so as it looks like a `torus'.
}
\label{Fig-schematic}
\end{figure}

The final point to note is that we can exclude the `thin'
almost-edge-on disk which was
proposed by \citet{Cuesta95}. This would give rise to a narrow, extinction
structure, with a strongly elongated `higher extinction region' perpendicular to the polar axis.
The observed structure is only mildly elongated, and the direction of
elongation is opposite to what would be expected from a thin disk. The
observed elongation is reminiscent of a cylindrical structure, or of a torus
with thickness comparable to its radius (Fig.~\ref{Fig-schematic}), but apart from the polar holes the
structure show no clear deviation from spherical symmetry.



An inevitable question is whether this dust shell is responsible for
collimating the bipolar flow.  In favour of this is the location, at the neck
of the flow, and the compact size, 10 per cent of that of the lobes.  The
similarity of dynamical age between the dust shell and the shell shows that
the collimation first occurred very early in the evolution, during the AGB or
in the early post-AGB wind.  This may argue against versions of the
Interacting Stellar Winds where the shaping is due to the fast wind from the
central star of the PN, as this wind only develops in the later post-AGB
evolution, at higher stellar temperature.

The timing of the lobes is consistent with current ideas that the
shaping originally occurs due to a change in the wind morphology at
the end of the high mass-loss phase \citep{Zijlstra01}. A model which
could fit the observed characteristics (a symmetric shell with lower
densities only towards the poles) is presented by \citet{Sahai98}:
they propose that a fast jet-like wind punches holes in the spherical
AGB shell, and that the subsequent post-AGB evolution amplifies this
sudden departure from spherical shape. 

Such a model may require a binary companion as source for the jet.
There is no evidence either against or in favour of a binary at the
core of NGC\,6537.  The optical magnitude rules out a solar-like star,
but a fainter red dwarf or a white dwarf are not excluded. The near
infrared detection may show some excess emission, but this cannot be
taken as evidence for a red companion. For comparison, the central
star of the very similar PN NGC\,6302 has been shown to have an
infrared excess which is not a stellar continuum \citep{Matsuura05}.

\subsection{ Hierarchical structure}

In NGC\,6537, a hierarchical structure is found, with the peak of PAH
intensity located further out than that of H$\alpha$.  This is consistent with
observations in other objects (NGC~7027 \citep{Woodward89} and NGC~6302
\citep{Matsuura05}), where PAH emission is located outside of the ionised gas.
The extinction continues to increase outside the PAH region, and PAH emission
is not found where the extinction is high ($\sim$5~arcsec).  This can be a
consequence of the lack of photo excitation in the high extinction region, or
a difference in abundances \citep{Matsuura04}.

The abundance analysis from optical line ratios shows that NGC~6537 is
oxygen-rich \citep{Pottasch00b}. The presence of PAHs \citep{Roche86}, which
are normally assumed to be formed from carbon-rich gas, is unexpected but this
chemical dichotomy is also seen in a few other PNe \citep{Matsuura05}.  PAHs
have smaller absorption coefficients in the optical than silicate grains, and
PAHs are unlikely to be the source of high extinction in the optical:
amorphous silicate is likely the main contributor to the optical extinction.
Crystalline silicates have also been detected in NGC~6537 \citep{Molster02}.
As the peak of extinction occurs outside of the PAH emission peak, the
silicate grains are located outside the location of the PAHs. This can
indicate that the silicates pre-date the PAHs.

Hydrogen molecules are detected in NGC\,6537: the brightest H$_2$ region is
found around the ionized core, corresponding to the high extinction shell
\citep{Davis03}. The presence of H$_2$ shows that the dust shell is partly
molecular, and the separation between extinction and H$\alpha$ traces an
ionization front. The PAH emission is located close to the ionization front
but the precise relation, and whether the PAHs formed in-situ or also date
from the AGB, remains to be determined. 

The hierarchy, with the PAH located in between the ionized and
molecular gas, suggests that the PAH emission may be associated with
the photo-dissociation region of the molecular dust shell.  Following
\citep{Matsuura04}, we associate the hot dust component (Section
\ref{dust}) with dust located in the PAH-emitting region.

\section{Conclusion}

The high-resolution images presented in this paper detect the central star of
NGC\,6537. The agreement between the optical and infrared images secure the
identification. The magnitudes are relatively uncertain, due to the high
background, but the derived values agree within the uncertainties with a hot
black body. The luminosity depends on the assumed temperature: dynamical
considerations, comparing with the stellar evolution tracks, argues for
relatively low luminosities $\sim 10^3\,\rm L_\odot$. This gives a temperature
in the range 1.5--2.5$\times 10^5\,\rm K$, slightly lower than the HeII
Zanstra temperature of $3.4 \times 10^5 \,\rm K$. We derive a core mass around
$M_c \approx 0.7$--$0.9\,\rm M_\odot$, and a progenitor mass $M_i = 3$--7\,M$_\odot$.

The progenitor mass is consistent with the classification of NGC\,6537 as a
type-I PN. Bipolar PNe have been suggested to have higher than average initial
mass: our results do not conflict with this.

The extinction map is not uniform inside NGC~6537, but shows a shell structure
of inner radius of 2\,arcsec and outer radius of at least 4\,arcsec.  The
central star is located at the centre of the near-symmetric extinction
distribution.  The dust shell traces the high mass-loss phase
during the AGB phase, and was expelled from the central star about 1400\,years
ago.  There is no evidence for a thin disk, as has been suggested.  The dust
shell is far more compact, and far more symmetric than the extended bipolar
lobes. A thick torus or cylindrical structure can fit the observed extinction
map.

The dynamical age of the lobes and dust shell are very similar. This
suggests that the origin of the bipolarity should be close to the end
of the AGB.  The largely symmetric dust shell, combined with the
strongly bipolar lobes, are consistent with models of a fast change of
wind structure at the end of the AGB mass loss. 

The structure shows radial hierarchy, with the ionized core located at the
inner edge of the dust shell, and the PAH emission in between, tentatively
associated with the photo-dissociation region.

\section{acknowledgements}

We are grateful for the support by the ESO staff during observations.
The ESO/ST-ECF Science Archive was used for this study.
M.M. is financially supported by PPARC.

\label{lastpage}

\end{document}